\documentclass[12pt]{iopart} 
\usepackage{iopams}   
\usepackage{graphicx,cite,epsf} 
 
\begin{document} 

\title[Directed and semiflexible polymers in anisotropic environment]{Lattice models of directed and semiflexible polymers in anisotropic environment} 

\author{K. Haydukivska} 
\address{Institute for Condensed Matter Physics of the National Academy of Sciences of Ukraine, 
79011 Lviv, Ukraine} 
\author{V. Blavatska} 
\address{Institute for Condensed Matter Physics of the National Academy of Sciences of Ukraine, 
79011 Lviv, Ukraine} 
\ead{viktoria@icmp.lviv.ua}

\begin{abstract} 

We study the conformational properties of polymers in presence of extended columnar defects of parallel orientation. 
Two classes of macromolecules are considered: the so-called partially directed polymers
 with preferred orientation along direction of the external stretching field
 and semiflexible polymers. We are working within the frames of lattice models: 
partially directed self-avoiding walks (PDSAWs) and biased self-avoiding walks (BSAWs).  
 Our numerical analysis of PDSAWs  reveals, that competition between
the stretching field and anisotropy caused by presence of extended defects leads to existing of 
three characteristic length scales in the system. At each fixed concentration of disorder we found 
 a transition point, where the influence of extended defects is exactly counterbalanced by the stretching field.
Numerical simulations of BSAWs in anisotropic environment reveal an increase of polymer stiffness. In particular, the  persistence length of semiflexible polymers increases in presence of disorder.

\end{abstract} 

{\it Keywords}: polymers, self-avoiding walks, computer simulations, conformational properties 

\submitto{\JPA}

\section{Introduction}

Conformational properties of  polymers are the subject of great interest during decades due to their important role both in nature and industry.
In statistical description of macromolecules it is established, that they are characterized by a number of properties, which are universal, i.e. independent on any details of microscopic chemical structure \cite{Gennes79,desCloizeaux90}. In particular, the mean-squared end-to-end distance $\langle R^2 \rangle$ of long flexible polymer chain in a good solvent scales with number of monomers (molecular weight) $N$ according to:
\begin{equation}
\langle R^2 \rangle \sim N^{2\nu}, \label{flex}
\end{equation}
here $\langle \ldots \rangle$ denotes averaging over an ensemble of conformations, and the universal size exponent $\nu=0.587597(7)$  in three dimensions \cite{Clisby00}.

Many synthetic polymers with a carbon backbone, such as polyethylene, can be considered as long flexible polymer chains: the molecular weight of repeating units (monomers) is rather small and
neighboring carbon atoms can easily rotate around single bonds.
However, not all polymers satisfy this criterion. In particular, the important organic macromolecules such as DNA and some proteins contain massive repeating units (e.g., radicals of
amino acids) with complicated interactions between them, which thus can not freely rotate on their bonds.
This class of polymers is usually referred to as semiflexible \cite{Bustamante94,Ober00}.
An important characteristic here is a persistence length $l_p$: when the total chain length is much larger
than $l_p$, any polymer behaves as a flexible chain obeying the scaling law (\ref{flex}), whereas for the chain
length much smaller than the persistence length, the polymer attains the limit of rigid rod with $\langle R^2 \rangle \sim N^{2}$.
 This type of ``coil-to-rod'' conformational  transition can be describe by presenting a size measure of polymer in a scaling form \cite{Halley85}:
\begin{equation}
\langle R^2 \rangle \sim N^{2} f\left(\frac{N}{l_p}\right).\label{f}
\end{equation}
Here, $f(x)$ is a scaling function:
\begin{equation}
f(x) = \left\{
  \begin{array}{l l}
    {\rm const},  & \quad N<<l_p, \\
    x^{2\nu-2} & \quad N>>l_p.
  \end{array} \right.
\end{equation}
Usually, the semiflexible polymers are characterized by a relatively
large value of persistence length $l_p$ (e.g., the mechanical persistence length of a
DNA macromolecule is of order of 50 nm \cite{Ober00}). Conformational properties of semiflexible polymers are intensively studied \cite{Camacho,Hsu12,Halley85,lee85,Moon91,Halley90,Hsu11coil,Hsu13,Stepanow08,Lam09,Privman87}.

Another interesting class of polymer macromolecules are so-called directed polymers \cite{Chakrabarti83}. Presence of anisotropy in the system, e.g. the external electrical field in the solution of polyelectrolytes, causes the stretching of polymer chain in direction parallel to the field, but allows fluctuations in transverse directions.  Due to anisotropy of directed polymers, one should distinguish between the transverse $\langle R_{\bot}^2 \rangle$ and longitudinal $\langle R_{\|}^2\rangle $ components of the end-to-end distance, which scale with the molecular weight $N$ according to:
\begin{equation}
\langle R_{\|}^2\rangle \sim N^{2\nu_{\|}}, \qquad \langle R_{\bot}^2 \rangle  \sim N^{2\nu_{\bot}}
\label{Rdir}
\end{equation}
where $\nu_{\|}=1$ and $\nu_{\bot}=1/2$ in all dimensions \cite{Cardy83}.

Universal scaling properties of polymers can be successfully described within the framework of lattice model of self-avoiding random walk (SAW) \cite{Vanderzande}. The fact, that the trajectory can not cross itself reflects the excluded volume effect in polymers. This model can be generalized  for the
case of semiflexible polymers by introducing different statistical weights for ``trans'' step
 (taken in the same direction as the previous one) and  ``gauche''  step (leading to a bending of the trajectory),
   which is known in literature as biased self-avoiding walks (BSAW) \cite{Glasser86,Halley85,lee85}. The persistence
length of BSAW trajectory is thus introduced as inversely proportional to the probability of a "gauche" step.
 The properties of  directed polymers can be described within  a model of directed random walk (DSAW).
 The DSAW trajectory is considered to be directed along some coordinate axis (say $x$), if projection of each step of the trajectory
 onto this axis is non-negative \cite{Chakrabarti83,Vanderzande}. One can treat this as a result of
 stretching field acting in direction $x$. 
 Note that the self-avoidance of directed walk trajectory (excluded volume effect of polymers)
 does not play any role, and  the size exponents in (\ref{Rdir}) are trivial in all dimensions \cite{Cardy83}.

In real physical systems, one often encounters various forms of structural inhomogeneities and irregularities. In studying the polymer systems, of great importance
is understanding of the behavior of macromolecules in the presence of  structural obstacles, e.g.  in colloidal 	solutions \cite{Pusey}, in gels \cite{Stylianopoulos10} or intra- and extracellular environment \cite{Verkman13,Xiao08}.  One can introduce disorder into the lattice model of polymer by considering some sites of the lattice as occupied by defects and forbidden for SAW trajectory.

{ In general, dealing with disordered systems one usually encounters two types of ensemble averaging,
treated as annealed and quenched disorder \cite{Brout59,Emery75}. 
In the former case, a characteristic time of impurities dynamics is comparable to relaxation times in the pure system, 
whereas in the latter case the impurities are
fixed and the configurational average over an ensemble of disordered systems with different
realization of the disorder is needed. 
Though in general the critical behavior of systems with quenched and
annealed disorder could be quite different,
 an equivalence of the influence of quenched and annealed disorder on scaling properties of long flexible polymer chains
 is proved \cite{Blav13}. 
}
Problems of influence of structural disorder on the conformational properties of polymer systems  are widely investigated over the decades. 
It was shown, that disorder in the form of uncorrelated point-like defects of weak concentration does not inflict the universal properties of polymers in good solution; in particular the  size exponent $\nu$  in (\ref{flex}) does not change its value. 
{ This statement has been proved by Harris \cite{Harris83}
and confirmed later both by renormalization group results \cite{Kim83} and numerical studies \cite{Kremmer81,Lee89,Lam90}.}
Only  when concentration of obstacles reaches a percolation threshold, the situation  changes drastically \cite{Blavatska08,Ordemann00,Janssen07,Grassberger93,Rintoul94,Lee96}:  { the scaling law (\ref{flex}) holds with exponent $\nu^{pc}=0.660(5)$ in $d=3$ \cite{Rintoul94}.} Similar situation can be found in the problem of  directed walks: some of the researches state that the value of transversal exponent $\nu^{pc}_{\bot}$ changes to $2/3$ \cite{Kardar87,Wang03} while some other appeals that it remains unchangeably $1/2$ \cite{santra01}.

Similarly as presence of external field can lead to spatial anisotropy and existence of two characteristic length scales of polymer macromolecule (\ref{Rdir}), presence of extended structural defects of parallel orientation (columnar defects)  also causes the analogous effect \cite{Baumgartner96}. 
In this case, one should again distinguish between the components of the length scale  in directions parallel  $\langle R_{||}^2 \rangle$ and perpendicular $\langle R_{\perp}^2\rangle $ to extended defects, which scale with nontrivial exponents $\nu_{\|}$ and $\nu_{\bot}$, respectively.
In Ref. \cite{Deutch05} it was shown that the inequality $\nu_{\bot}<\nu<\nu_{\|}$ should hold, where $\nu$ is a size exponent of polymer in pure environment (\ref{flex}). 
The conformational properties of flexible polymer chains in presence of extended defects in form of lines of parallel orientation  have been studied in Refs. \cite{Deutch05,Haydukivska14} and  numerical estimates for exponents $\nu_{\bot}$, $\nu_{\|}$ obtained. On the other hand, it was shown, that the presence of extended columnar defects does not alter the universal properties of directed polymers \cite{Arsenin94}.

In present work, we study the conformational properties of partially directed and semiflexible polymers in anisotropic environment with extended columnar defects. The layout of the paper is as follows. In the next Section, we introduce the lattice models of partially directed and biased self-avoiding walks. In Section 3, we shortly described the Pruned-Enriched Rosenbluth method, which is applied for analysis of properties of the models in Section 4.
We end up with conclusions in Section 5.   

\section{The Model}
\label{model}

\begin{figure}[t!]
\begin{center}
\includegraphics[width=40mm]{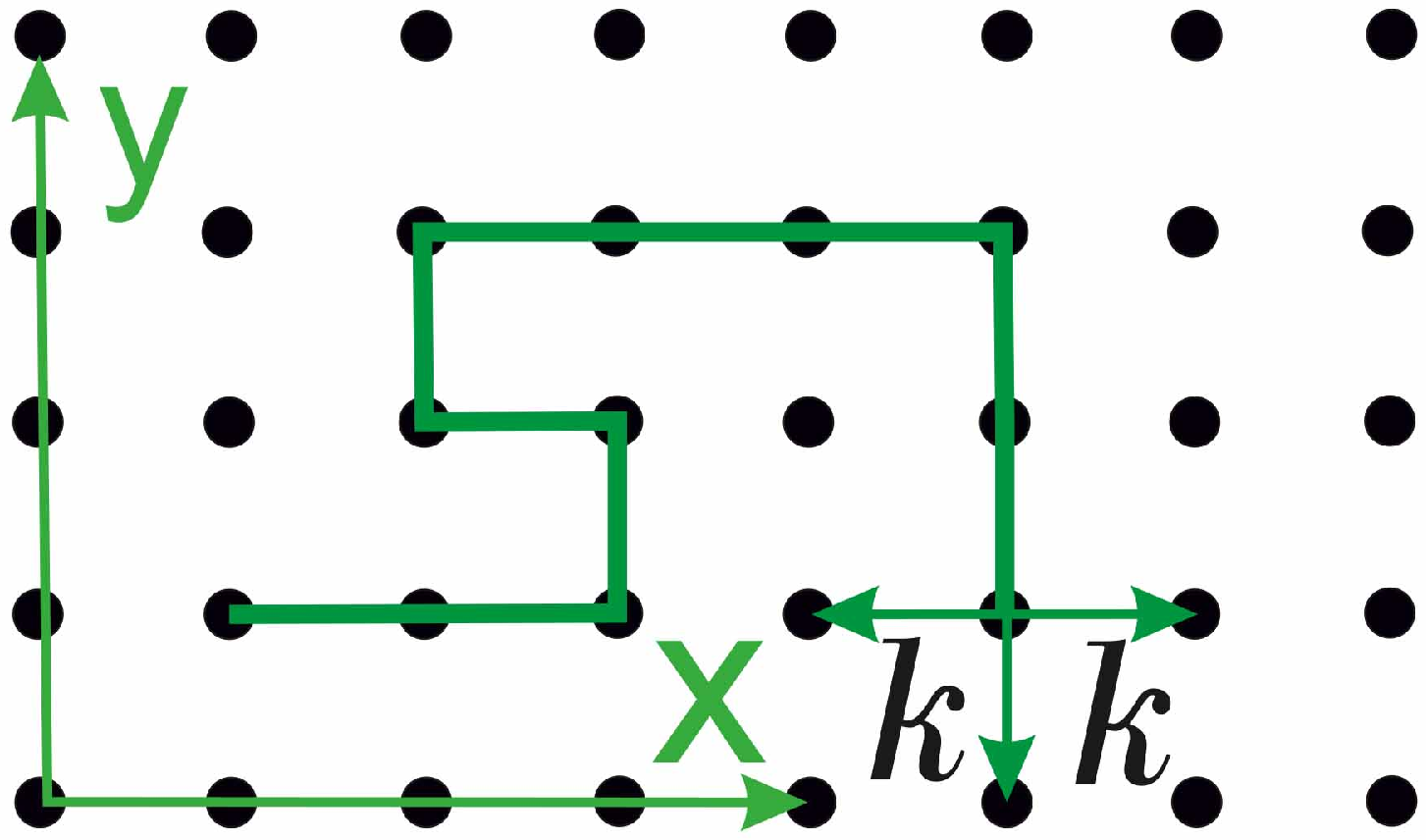}
\hspace*{1cm}
\includegraphics[width=40mm]{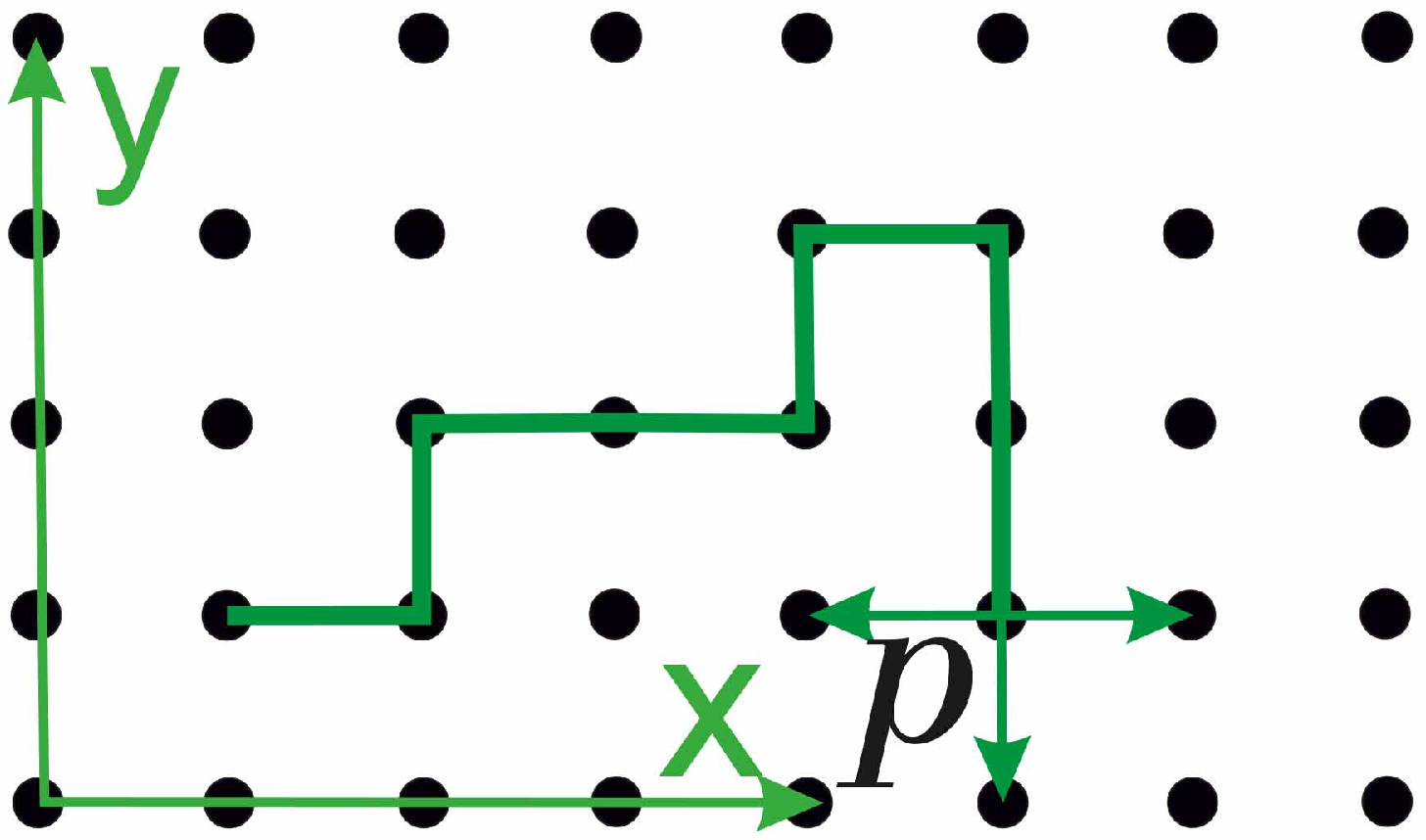}
\end{center}
\caption{Left: Schematic presentation of semiflexible polymer within the framework of BSAW model. The ``gauche'' steps leading to a bending of the trajectory are allowed with probability $1/3\cdot(1-k)$. Right: Schematic presentation of partially directed polymer within the framework of PDSAW model. The step with negative projection in direction $x$ is allowed with probability $1/3\cdot(1-p)$. }
\label{fig:2}
\end{figure}

To study the universal conformational properties of  semiflexible and directed polymers in anisotropic environment, we start with the simple model of self-avoiding random walks on a regular lattice.
We are working within the frames of Rosenbluth growing chain algorithm \cite{Rosenbluth55}: SAW trajectory grows step by step until it reaches the total length $N$. For each $n$th step ($n<N$) of growing trajectory, there are
$z_n$ available nearest neighbor sites to make $n+1$th step, which variates from $0$ to $2d-1$ due to requirement that the SAW cannot cross itself and visit the same site more than once. If $z_n=0$ at $n<N$,  the process is stopped and a new chain is growing, until the desired total length $N$ is reached. Thus, $n+1$th step is performed towards any of $z_n$ directions with equal probabilities
 $\rho_n=1/z_n$.
 To generalize this model  for the
case of semiflexible polymers,  we take into account that  steps in directions, which lead to bending of the trajectory (see Fig. \ref{fig:2} left) are made with smaller probabilities $\rho_n \cdot (1-k)$
 with stiffness parameter $0\leq k \leq 1$ \cite{Halley85}.  This corresponds to  biased self-avoiding walks (BSAW).  With $k=0$ one restores the simple SAW (flexible polymer chain), whereas $k=1$ corresponds to the limit of completely stiff, rod-like polymer structures.
 To study the properties of partially directed polymers under the acting of stretching field, we take into account that
 the step with negative projection in direction $x$  (see Fig. \ref{fig:2} right) could be performed with smaller probability $\rho_n \cdot (1-p)$ with stretching parameter $0\leq p \leq 1$. This corresponds to partially directed self-avoiding walks (PDSAW).
By changing $p$ from $0$ (limit of flexible chain -- SAW) to $1$ (directed polymer chain -- DSAW) we can study the properties of conformational transitions between these states.

\begin{figure}[b!]
\begin{center}
\includegraphics[width=50mm]{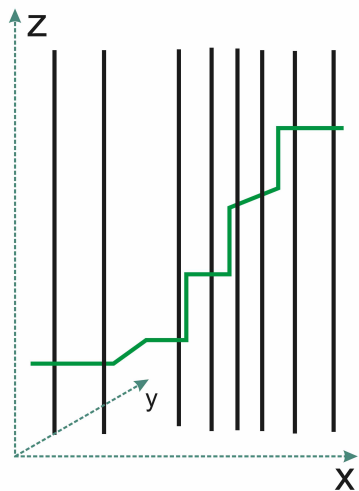}
\end{center}
\caption{Schematic presentation of directed polymer chain in anisotropic environment.  The longitudinal direction of polymers ($x$) and direction of alignment of extended defects ($z$) are perpendicular to each other.}
\label{fig:1}
\end{figure}

{
We are interested to analyze the properties of these models in presence of quenched disorder in the form of parallel lines, stretching out along direction $z$ 
and randomly distributed in $xy$-plane of a lattice. Thus, the ``projections'' of these lines onto the $xy$ plane of the lattice 
form a set of
point-like uncorrelated defects.     
To realize this type of disorder, we start with $xy$-plane of a pure lattice. 
Each site on this plane is then assigned to be occupied by point-like defect (and thus not allowed for SAW trajectory)
with given probability $c$  and empty otherwise.
Then, we built the lines in direction $z$ through
each site containing point-like defects   (see Fig. 2). 
}
We consider concentrations of disorder  up to $40$ percents due to the fact that on $d=2$-dimensional lattice there is a percolation threshold closely to this concentration, and disorder with $c>0.4$ leads to trivial results \cite{Haydukivska14}.

\section{The Method}

The Pruned-Enrichment Rosenbluth Method (PERM) \cite{Grassberger97} is used to explore the peculiarities of the model. The PERM is based on original algorithm of growing chain proposed by Rosenbluth and Rosenbluth \cite{Rosenbluth55} with additional population control parameters. On every $n$-th ($n\leq N$) step of growing SAW trajectory
 a weight $W_n$ is assigned to a chain:
\begin{equation}
W_n= \prod^n_{i=1} m_i
\label{W}
\end{equation}
where $m_i$ is a number of available nearest neighbor sites. 
Note that in calculating $m_i$ one should take into account, that in the case of BSAW or PDSAW the steps in directions, which lead to bending of the trajectory or steps with negative projection on axis $x$, respectively, should be properly taken with smaller probabilities (see previous section), which leads in general to decreasing of $m_i$ as comparing with $z_i$.
Additionally, it is important to note that the growing trajectory cannot visit lattice sites occupied by defects (in our case,  these sites form a set of parallel lines that are randomly distributed in $xy$ plane and penetrate through the whole system).

When the chain of total length $N$  is constructed, the new one starts from the same starting point, until the desired number of chain configurations are obtained.
{ Note that we are dealing with the case of quenched disorder, 
 performing the  configurational average  of any observable 
 over an ensemble of  different
realizations of disordered lattices according to:}
\begin{equation}
\langle (\ldots)
\rangle=\frac{1}{Z_N}{\sum_{k=1}^{M}W_N^{k}(\ldots)},
\,\,\,\,Z_N=\sum_{k=1}^{M} W_N^{k}, \label{conn}
\end{equation}
here $W_N^{k}$ is a weight of $k$-th configuration of $N$-step trajectory and $M$ is number of configurations.
Though in general the critical behavior of systems with quenched and
annealed disorder could be quite different,
 an equivalence of the influence of quenched and annealed disorder on scaling properties of long flexible polymer chains
 is proved \cite{Blav13}.

While the chain grows by adding monomers, its weight fluctuates. PERM suppresses these fluctuations by pruning configurations with too small weights, and by enriching the sample with copies of high-weight configurations
\cite{Grassberger97,Hsu03}. Pruning and enrichment are performed by choosing thresholds $W_n^{<}$
and $W_n^{>}$, which are continuously updated as the simulation progresses. We take \cite{Hsu03}: $W_n^{>}=C(Z_n/Z_1)(c_n/c_1)^2$ and
$W_n^{<}=0.2W_n^{>}$, where $c_n$ is the number of created chains of length $n$, and the parameter $C$ controls the pruning-enrichment
statistics; it is chosen in such way that on average 10 chains of total length $N$ are generated per each tour.

\section{Results and discussions}

In the present work, we study  conformational properties of  partially directed and semiflexible polymers in anisotropic environment with quenched disorder in form of extended columnar defects.
It is important to note that in our problem we have to perform two types of averaging of any observable: the first average is performed over all BSAW or PDSAW configurations on a lattice with fixed configuration of impurities according to (\ref{conn});
the second average is carried out over different realizations (replicas) of disorder:
\begin{equation}
\overline {(\ldots)}=\frac{1}{M}{\sum_{i=1}^{M}(\ldots)},\label{average}
\end{equation}
where $M$ is a number of replicas.
We use cubic lattices of the size $L=300$ and performed averages over
 $400$ replicas with a set of up to $10^5$ chains on each of them.

\subsection{Partially directed polymers}

We analyze the properties of partially directed polymers in anisotropic environment within the frame of PDSAW model under varying both the stretching parameter $p$ and concentration of columnar defects $c$.

\begin{figure}[t!]
\begin{center}
\includegraphics[width=90mm]{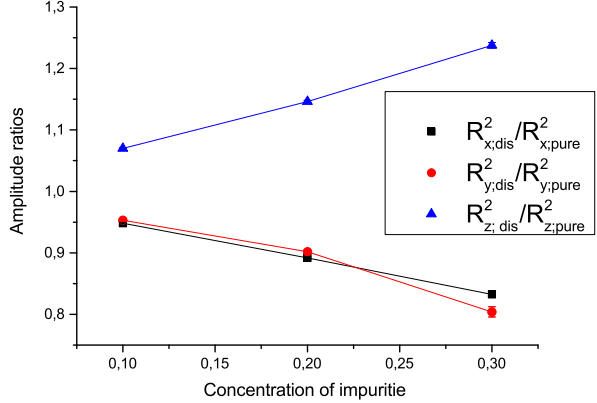}
\end{center}
\caption{Amplitude ratios as functions of concentration of disorder.}
\label{fig:6}
\end{figure}

Let us recall, that in our model the polymers are considered to be directed along the $x$-axis, whereas 
the extended columnar defects are oriented along $z$-axis (Fig. \ref{fig:1}), so that the stretching field acts in direction
 perpendicular to the lines of defects.
Thus, we have three characteristic directions in the systems, and corresponding components of 
the end-to-end distance of PDSAW are expected to behave differently:
\begin{eqnarray}
&&\overline{\langle R_{x}^2 \rangle} =(x_N-x_0)^2\sim N^{2 \nu_{x}}, \nonumber\\
&&\overline{\langle R_{y}^2 \rangle} =(y_N-y_0)^2\sim N^{2 \nu_{y}}, \nonumber\\
&&\overline{\langle R_{z}^2\rangle} =(z_N-z_0)^2\sim N^{2 \nu_{z}}\label{R1}
\end{eqnarray}
with non-trivial size exponents $\nu_{x}$, $\nu_{y}$, $\nu_{z}$.  In case when no disorder is present in the system, we restore $\nu_{x}=\nu_{||}$, $\nu_{y}=\nu_{z}=\nu_{\perp}$ (cf. (\ref{Rdir})).

\begin{figure}
\begin{center}
\includegraphics[width=90mm]{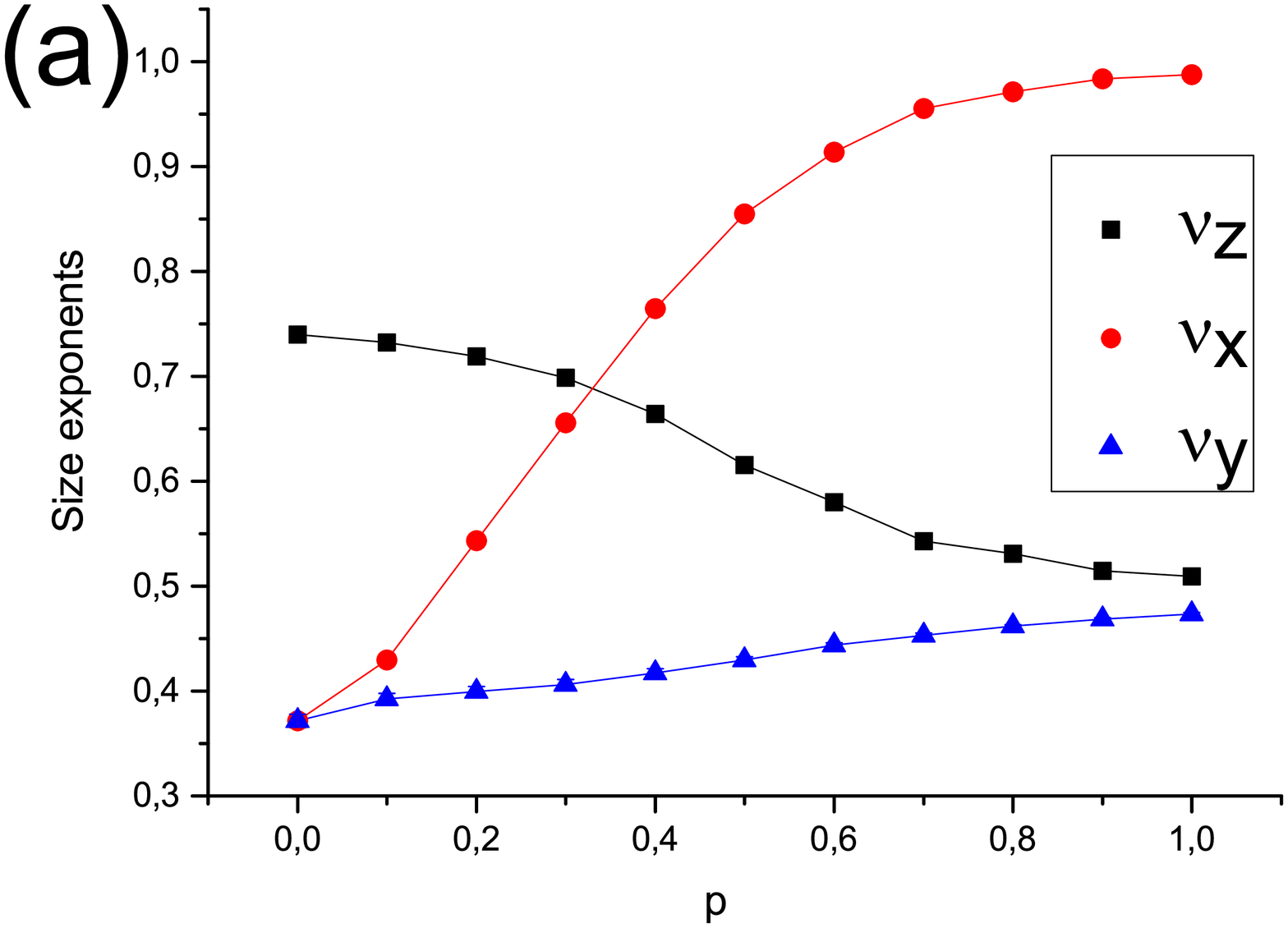} 
\includegraphics[width=100mm]{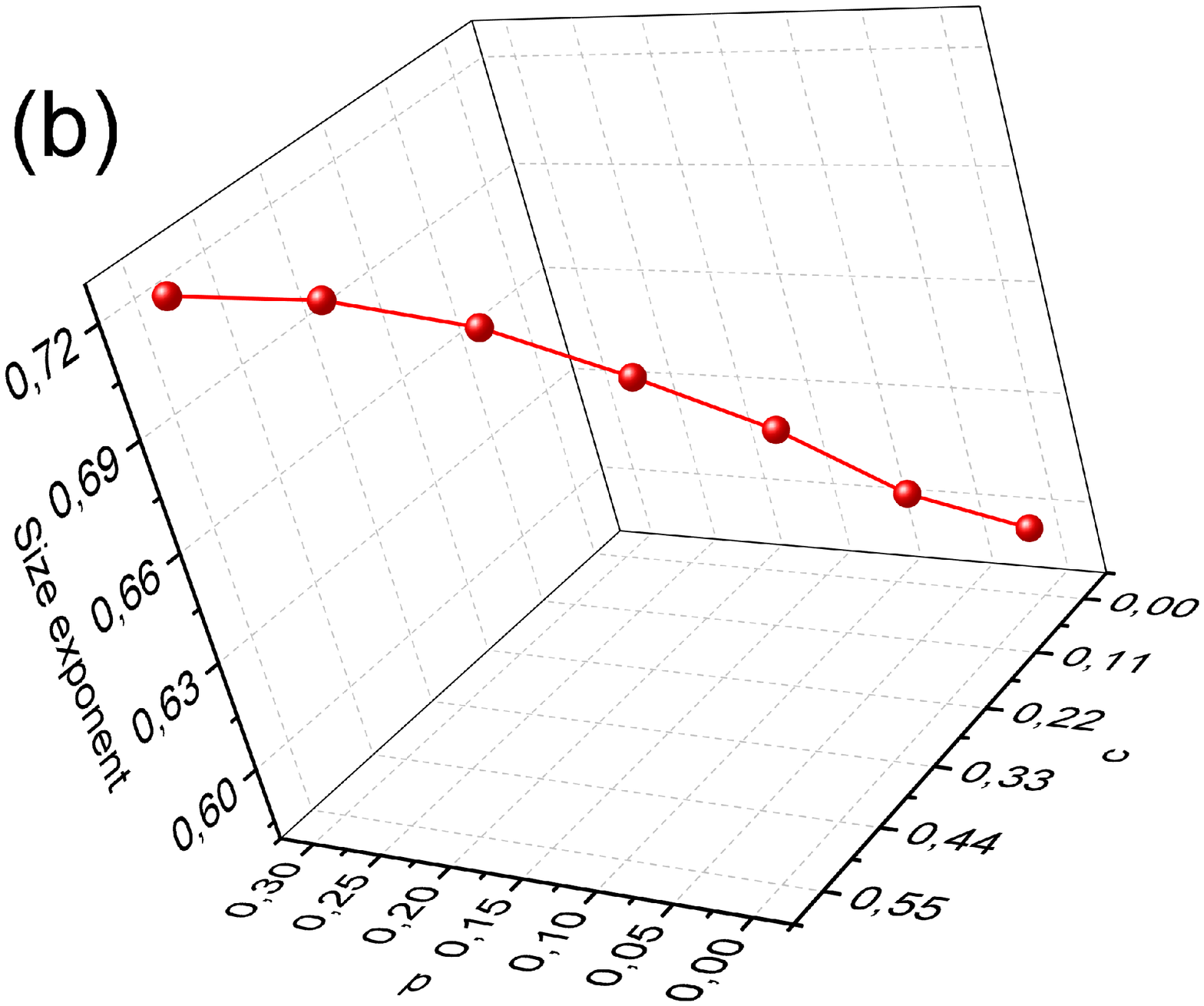}
\end{center}
\caption{(a): Size exponents of partially directed polymers as functions of stretching parameter $p$ at fixed concentration of linear defects $c=0.2$. (b): Size exponent $\nu_{xz}$ as function of stretching parameter $p$ and concentration of linear defects $c$.}
\label{fig:7}
\end{figure}

We start with calculation of the size exponents for completely directed polymers ($p=1$). The data obtained lead us to  conclusion that, in accordance with Ref. \cite{Arsenin94}, influence of columnar disorder is rather trivial. The values of size exponents (\ref{R1}) governing the scaling of components of $\overline{\langle {R}_{{\rm dis}}^2\rangle}$ in presence of disorder are not modified as compared with those of DSAW on pure lattices (${\langle {R}_{{\rm pure}}^2\rangle}$). Thus, the anisotropy caused by stretching field plays the dominant role comparing with an effect of anisotropy caused by disorder.
At this point it is interesting to consider the amplitude ratios $\frac{\overline{\langle R_{x,{\rm dis}}^2\rangle}}{{\langle R_{x,{\rm pure}}^2 \rangle}}$,
  $\frac{\overline{\langle R_{y,{\rm dis}}^2\rangle}}{{\langle R_{y,{\rm pure}}^2\rangle}}$, 
  $\frac{\overline{\langle R_{z,{\rm dis}}^2\rangle}}{{\langle R_{z,{\rm pure}}^2\rangle}}$,
which are plotted on Fig. \ref{fig:6} as functions of concentration of disorder $c$.
These ratios may help us to catch the effects, caused by presence of extended impurities,  
even when the universality class remains unchanged. 
One notes that presence of  parallel columnar defects 
leads to slightly shrinking of polymer in directions $x$ and $y$  (decreasing of corresponding amplitude ratios) 
and correspondingly to slight expansion in direction $z$.

To find the numerical estimates for the critical exponents in (\ref{R1}), we analyzed the behavior of corresponding components of the end-to-end distance as functions of the length of growing PDSAW trajectory under varying both the stretching parameter $p$ and concentration of defects $c$ and applied the least-square fitting to the data obtained. The results obtained are presented on Fig. \ref{fig:7}a. 
{ Let us recall, that the critical exponents of simple SAW are universal in the sense that they are dependent on the only one ``global parameter'' -- the space dimension $d$ (see e.g. Eq. (\ref{flex})). In our case of PDSAW, we note the non-universal behavior of exponents $\nu$ as  functions of the stretching parameter $p$, which now plays the role of another global parameter. Note, that similar non-universality is obtained in renormalization group studies of BSAW model in Ref. \cite{Giacometti92},
 where, in particular, the dependence of critical exponent $\nu$ on the stiffness parameter (bending energy) is obtained.
}
If the stretching field is absent ($p=0$), we restore the problem of simple SAW in presence of linear defects studied in \cite{Haydukivska14}. In this case, as expected, one finds stretching of trajectory in direction, parallel do orientation of defects, so that 
$\nu_{z}>\nu_{x}=\nu_{y}$.  On the other hand, the competition between
the stretching field and space anisotropy caused by presence of defects leads to existing of three characteristic length scales in the system.    
With increasing the parameter $p$ at fixed concentration of disorder,
the stretching of polymer in direction $x$ increases ($\nu_{z}>\nu_{x}>\nu_{y}$) and  finally overcomes the effect caused by disorder ($\nu_{x}>\nu_z>\nu_{y}$ for large values of $p$).
Thus, at each fixed concentration of disorder, there should exist some critical value of stretching parameter $p$,
when the anisotropy caused by presence of extended defects is exactly ``counterbalanced'' by the stretching field, acting in perpendicular direction, so that $\nu_x=\nu_z\equiv\nu_{xz}$. 
This  may be treated as a transition point between the state where the role of disorder is dominant and the polymer is mainly oriented along $z$ axis, and the state when the stretching field starts to dominate and causes the polymer chain reorientation mainly along $x$ axis. 
Analyzing our data under varying both the stretching parameter $p$ and concentration of columnar defects $c$,
we found the pairs of $c$ and $p$ values corresponding to counterbalance between two  types of anisotropy, and estimated the values of exponents $\nu_{xz}$, presented  on Fig. \ref{fig:7}b.

\subsection{Semiflexible polymers}

\begin{figure}[b!]
\begin{center}
\includegraphics[width=90mm]{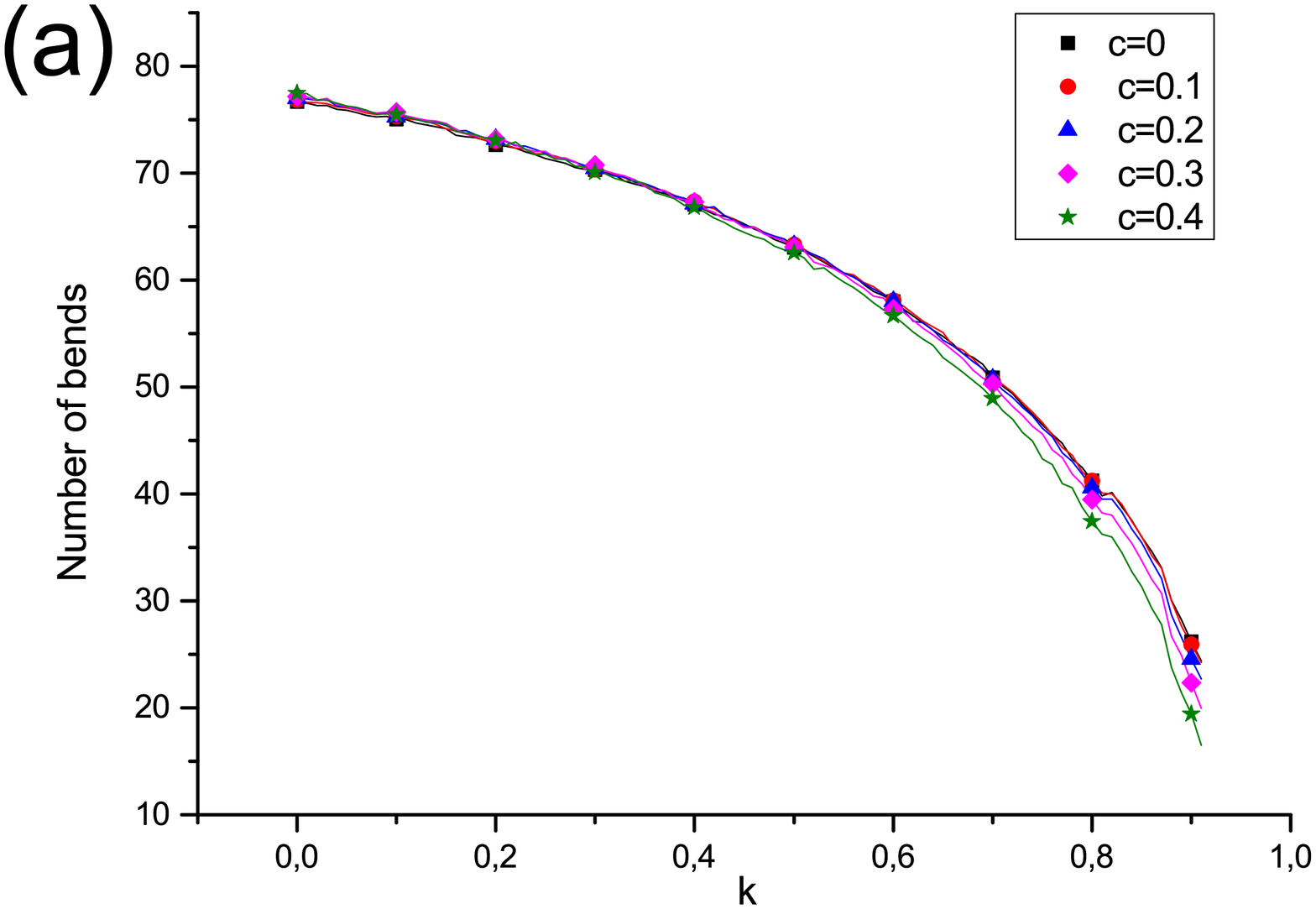}
\includegraphics[width=90mm]{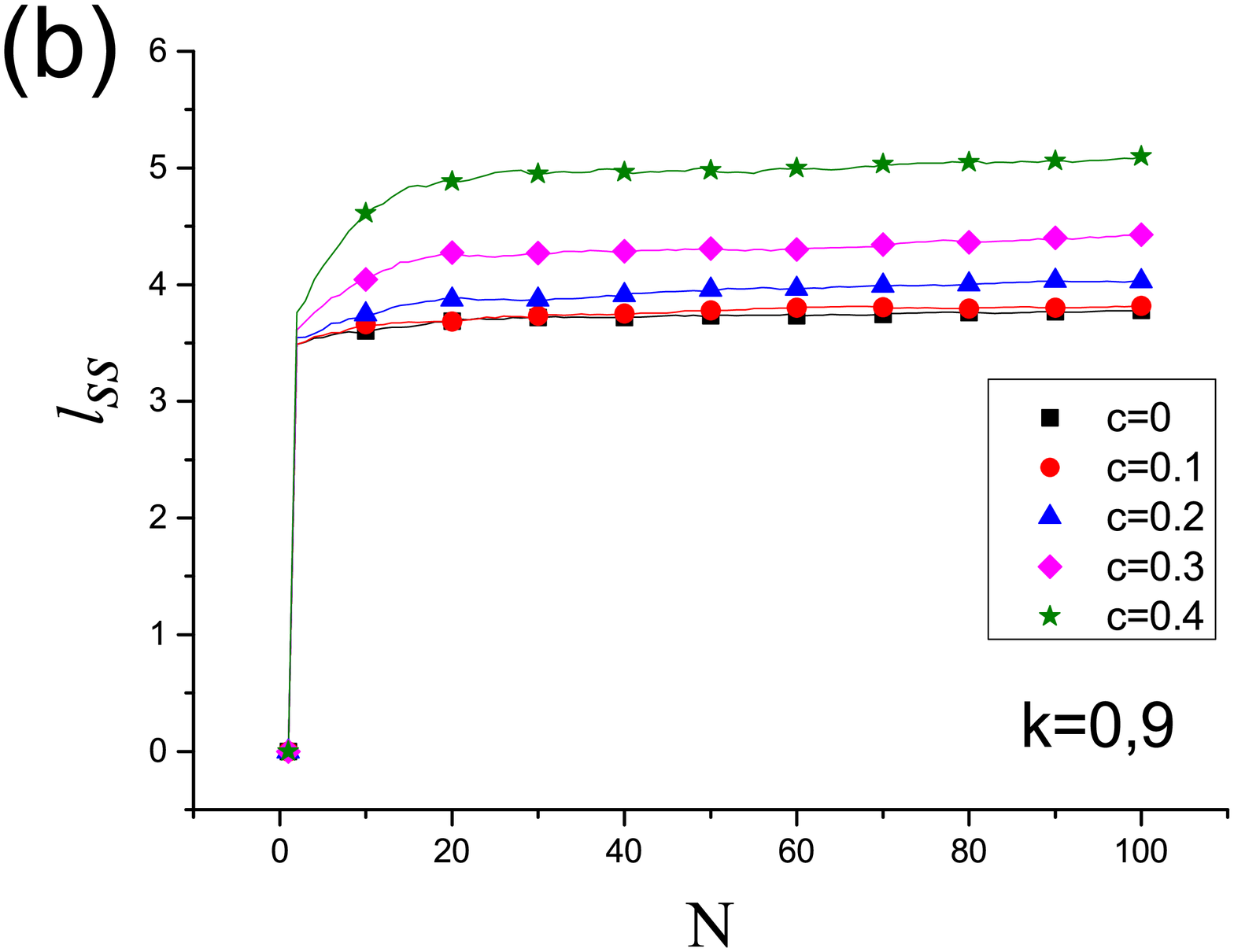}
\end{center}
\caption{(a): Number of bends in BSAW trajectory of total length $N=100$ at different concentrations of disorder $c$ as function of stiffness parameter $k$. (b): Average length of straight segment as function of chain length for different concentrations}
\label{fig:3}
\end{figure}

We analyze the properties of semiflexible polymers in anisotropic environment within the frames of
 BSAW model under varying both the stiffness parameter $k$ and concentration of columnar defects $c$.
 Numerical simulation are performed for the chain lengths up to $100$ monomers.

\begin{figure}[t!]
\begin{center}
\includegraphics[width=90mm]{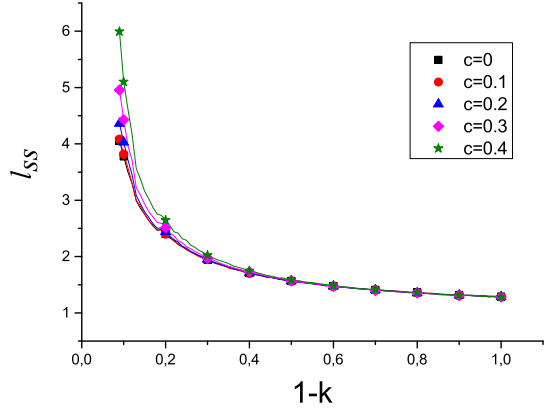}
\end{center}
\caption{Average length of straight segment as a function of stretching parameter for different concentrations}
\label{fig:12}
\end{figure}

We start with evaluating the number of bends $n_b$ in a polymer chain of fixed length $N$,  corresponding to a number of times when BSAW trajectory changes its direction, as function of parameters $k$ and $c$ (Fig. \ref{fig:3}a). As expected, $n_b$ decreases gradually with increasing the stiffness of a chain.
At small values of $k$, corresponding to regime of flexible polymer chain, presence of columnar defects
does not alter the bending of trajectory, whereas for considerably stiff chains (with $k$ close to $1$) the number of bends slightly decreases with increasing the concentration of disorder.
This effect can be visualized more clearly by analyzing the average number of steps between two successive bends
in BSAW trajectory (average length of a straight segment $l_{ss}$, which can be related to the  persistence length of semiflexible polymer), which increases with  disorder concentration (see fig. \ref{fig:3}b). 
For fixed chain length we can analyze the behavior of $l_{ss}$ as function of $1-k$ (see fig. \ref{fig:12}). These numerical data can be approximated according to:
\begin{equation}
l_{ss} = a(1-k)^b,\label{lss}
\end{equation}
where the values of constants $a$ and $b$ depend on the concentration of columnar disorder (see Table \ref{con}).

\begin{table}[b!]
\begin{center}
  \begin{tabular}{| c | c | c |}
      \hline
    $c$ & $a$ & $b$ \\ \hline
    $0$&$ 1.234(6) $&$ -0.382(5)$\\ \hline
    $0.1$&$1.229(6) $&$-0.385(6) $\\ \hline
    $0.2$&$1.224(6) $&$-0.394(6) $\\ \hline
    $0.3$&$1.217(7)  $&$-0.411(6) $\\ \hline
    $0.4$&$ 1.205(8) $&$-0.444(8) $\\ \hline
    \end{tabular}
\end{center}
\caption{Values of parameters $a$ and $b$ in Eq. (\ref{lss}) at different concentrations of disorder $c$.} 
\label{con}
\end{table}

\begin{figure}
\begin{center}
\includegraphics[width=90mm]{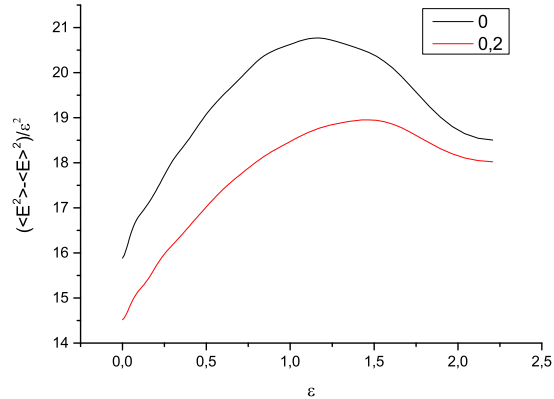}
\end{center}
\caption{Specific heat $C_N$ as function of bending energy at different concentrations of disorder.}
\label{fig:4}
\end{figure}

Note, that one can introduce a bending energy $\varepsilon$ associated to each ``gauche'' step, so that
the statistical weight ${\rm e}^{-\varepsilon}$ corresponds to each bend of the BSAW trajectory. With $\varepsilon=0$
 we restore the flexible polymer chain ($k=0$), whereas with increasing $\varepsilon$ the bends become energetically
 unfavorable,  so that $\varepsilon\to\infty$ corresponds to limit of stiff rod-like polymers ($k=1$).
One thus has \cite{Halley85}:
\begin{equation}
1-k=e^{-\varepsilon}.
\end{equation}
The total energy of $N$-monomer semiflexible polymer chain is thus given by $E_N=n_b\cdot \varepsilon$.
Statistical fluctuations of the energy $E_N$ indicates transitions or crossovers between different conformational states. In the case of semiflexible polymers, this corresponds to the transition between  regimes of flexible and stiff chains. We considered generalized ``specific heat'' per one monomer $C_N$ defined via energy fluctuations as:
 \begin{equation}
C_N=\frac{1}{N\varepsilon^2}\left(\overline{\langle E_N^2 \rangle}- \overline{\langle E_N \rangle^2}\right),
\label{cvperc}
\end{equation}
and analyzed  the peak structure of $C_N$ as function of $\varepsilon$ at various concentration $c$
of linear defects (Fig. \ref{fig:4}).
 In the case of pure environment, we find for the critical value of bending energy $\varepsilon\approx1.47$ 
 (corresponding to $k=0.77$). This value is shifted with increasing the concentration of disorder (at $c=0.2$ we observe $\varepsilon\approx 1.24$ $(k\approx0.71)$. Thus, as expected, the presence of 
  extended impurities makes the polymer chain stiffer and causes the conformational transition into 
  the rod-like state to take place at smaller values of bending energy $\varepsilon$ comparing with the case of pure environment.

\section{Conclusions}

The aim of the present paper was to analyze the conformational transformations in polymers in solutions 
in presence of columnar obstacles of parallel orientation, causing the spatial anisotropy. 
We consider two classes of polymers: the so-called partially directed  polymers with 
preferred orientation along direction of the external stretching field
 and semiflexible polymers.
 
 We are working within the frames of model of self-avoiding random walks (SAWs) on a regular lattice.
 To study the properties of partially directed polymers, we introduce the
  stretching parameter $0\leq p \leq 1$, and by changing $p$ from $0$ (limit of flexible chain -- SAW) to $1$ (directed polymer chain -- DSAW) we analyze the conformational transitions between these states.
This corresponds to partially directed self-avoiding walks (PDSAW).
 To generalize the SAW model for the
case of semiflexible polymers, we introduce the stiffness parameter $0\leq k \leq 1$: with $k=0$ one restores the simple SAW (flexible polymer chain), whereas $k=1$ corresponds to the limit of completely stiff, rod-like polymer structures. This corresponds to biased self-avoiding walks (BSAW).  

 Our numerical analysis of PDSAW model in an anisotropic environment with extended columnar defects reveals a new universality class with non-trivial values of critical exponents. The competition between
the stretching field and space anisotropy caused by presence of defects leads to existing of three characteristic length scales in the system. At each fixed concentration of disorder $c$, we found the critical value of stretching parameter $p$, when the anisotropy caused by presence of extended defects is exactly ``counterbalanced'' by the stretching field, acting in perpendicular direction. 
This  may be treated as a transition point from the conformation when the polymer is mainly oriented along the position of extended linear defects to the regime  when the stretching field starts to dominate and make the polymer chain reorientation mainly along the perpendicular direction. 

Numerical simulations of semiflexible polymers in anisotropic environment also reveal difference in comparison with pure environment. As expected, presence of extended columnar defects of parallel orientation leads to an increase of the stiffness of a chain. In particular, the average number of steps between two successive bends
in BSAW trajectory (average length of a straight segment $l_{ss}$, which is related to the persistence length of semiflexible polymer) increases with disorder concentration.
Analysis of the  peak structure of specific heat shows the shift of bending energy value with increasing the concentration of disorder and thus,  the presence of 
  extended impurities  causes the conformational transition into 
  the rod-like state at smaller values of $\varepsilon$ comparing with the case of pure environment.

\end{document}